\documentclass{aa}  

\newcommand{\gta}{\lower 0.5ex\hbox{$ \buildrel>\over\sim\ $}}
\newcommand{\lta}{\lower 0.5ex\hbox{$ \buildrel<\over\sim\ $}}

\newcommand{\teff}{$T_{\rm eff}$}
\newcommand{\Msun}{$M_{\rm \odot}$}

\usepackage{natbib}
\usepackage{graphicx}
\usepackage{txfonts}
\bibpunct{(}{)}{;}{a}{}{,}

\begin{document}

\title{Rotational broadening and conservation of angular momentum
  in post-extreme horizontal branch stars}

\author{G. Fontaine\inst{1} and M. Latour\inst{2}}  

\institute{D\'epartement de Physique, Universit\'e
  de Montr\'eal, Succ. Centre-Ville, C.P. 6128, Montr\'eal, QC, H3C 3J7, Canada \\
  \email{fontaine@astro.umontreal.ca } 
\and
  Dr. Karl Remeis-Observatory \& ECAP, Astronomical
  Institute, Friedrich-Alexander University Erlangen-Nuremberg,
  Sternwartstr.~7, 96049 Bamberg, Germany \\
  \email{marilyn.latour@fau.de} }

\date{Received ; accepted }  
  
\abstract{
We show that the recent realization that isolated 
post-extreme horizontal branch (post-EHB) stars are generally
characterized by rotational broadening with values of $V_{\rm rot} \sin
i$ between 25 and 30 km~s$^{-1}$ can be explained as a natural
consequence of the conservation of angular momentum from the previous
He-core burning phase on the EHB. The progenitors of these evolved
objects, the EHB stars, are known to be slow rotators with an average
value of $V_{\rm rot} \sin i$ of $\sim$7.7 km~s$^{-1}$. This implies
significant spin-up between the EHB and post-EHB phases. Using
representative evolutionary models of hot subdwarf stars, we demonstrate
that angular momentum conservation in uniformly rotating strutures
(rigid-body rotation) boosts that value of the projected equatorial
rotation speed by a factor 
$\sim$3.6 by the time the model has reached the region of the surface
gravity-effective temperature plane where the newly-studied post-EHB
objects are found. This is exactly what is needed to account for their
observed atmospheric broadening. We note that the decrease of the moment
of inertia causing the spin-up is mostly due to the redistribution of
matter that produces more centrally-condensed structures in the post-EHB
phase of evolution, not to the decrease of the radius per se.
}

  
 \authorrunning{Fontaine \& Latour}
 \titlerunning{Rotational broadening and conservation of angular momentum
  in post-EHB stars}

\maketitle

\section{Astrophysical context}

The hot, hydrogen-rich subdwarf O (H-sdO) stars are believed to be the
direct descendants of the cooler He-core burning stars of spectral type
sdB that populate the EHB. They correspond to the He-shell burning phase
(usually referred to as the post-EHB phase) that immediately follows the
EHB stage \citep{dor93,han03}. They are found in a
very wide range of effective temperature (\teff), from about 38\,000 K to
upward of 80\,000 K, and their surface gravities cover the interval 4.6
$\la$ log $g$ $\la$ 6.4 (see, e.g., Heber \citeyear{heb09,heb16}). While the main
atmospheric properties of these hot stars (\teff, log $g$, and log $N$(He)/$N$(H)) can
generally be estimated from optical spectroscopy coupled to appropriate
NLTE atmosphere models \citep{stro07,nem12,font14}, very few of them have been characterized in
terms of their chemical composition in detail. This is due both to the
scarcity of metal lines in the optical spectra of sdO stars along with
the inherently demanding task of computing NLTE models with metal line
blanketing. 

In this context, the recent study of Latour et al. (2017) has added four
stars to the meager sample of three H-sdOs for which detailed
abundance analyses using multiwavelength data have been carried out. 
The latter are \object{AA Dor} \citep{fle08,klepp11},
\object{BD$+$28$\degr$4211} \citep{lat13}, and \object{Feige 110} \citep{rauch14}. 
The additions of \citet{lat17} are \object{Feige 34}, \object{Feige 67},
\object{AGK$+$81$\degr$266}, and \object{LS II$+$18$\degr$9}, which have been found to
show very similar abundance patterns and to be located very close to
each other, around \teff $\simeq$ 61\,400~K and log $g$ $\simeq$ 6.0
(their average values) in the surface gravity-effective temperature
plane. Among other results, and of particular interest here, all of
these four stars have spectra that show the signature of rotational
broadening with characteristic values of $V_{\rm rot} \sin i$ between 25
and 30 km~s$^{-1}$. This adds significantly to the very few cases
reported so far in which rotational broadening could be detected in
H-sdOs --- essentially on the basis of rare high-resolution UV
observations. 

For instance, \citet{del92} were the first to report the need
for a broadening mechanism with an amplitude of some 30 to 40 km
s$^{-1}$ in the \textit{IUE} spectra of both LS II$+$18$\degr$9 
 and \object{CPD$-$71$\degr$172B}. The latter is a near
spectroscopic twin of the former, although CPD$-$71$\degr$172B resides in a binary
system with an F3$-$F4 main-sequence star \citep{vit88}.
For their part, \citet{beck95a} used a rotational
velocity of 30 km~s$^{-1}$ to account for the profiles of the metal
lines detected in the \textit{IUE} spectrum of Feige 67. More recently, \citet{rin12} best
reproduced the \textit{FUSE} spectrum of the H-sdO star \object{EC~11481-2303} (\teff =
55\,000 $\pm$ 5000 K, log $g$ = 5.8 $\pm$ 0.3) by using a value of
$V_{\rm rot} \sin i$ = 30 km~s$^{-1}$ as well. Put together with the
work of \citet{lat17}, this suggests that rotational broadening
of that order is likely a general characteristic of isolated post-EHB
stars.\footnote{We note, in this context, that the rotational velocity
  of $V_{\rm rot}$ = 35 $\pm$ 5 km~s$^{-1}$ \citep{fle08} or 30 km
  s$^{-1}$ \citep{klepp11} inferred for the H-sdO primary of the
  close binary system AA Dor, comparable in magnitude to the above
  values, cannot be considered representative of isolated stars,
  however. In comparison, the isolated well-studied star Feige 110 does
  not show obvious rotational broadening \citep{rauch14}, but it may
  be a genuine slow rotator or it could be seen under an unfavorable
  angle of inclination. As for the other well-studied very hot H-sdO
  star, BD$+$28$\degr$4211, the absence of detectable rotational
  broadening could also be due to the fact that this object may well be a
  post-AGB star, not a post-EHB star \citep{her99,lat13}.} 

In comparison, it has been shown that the immediate progenitors of H-sdOs,
the EHB stars, are collectively much slower rotators. Indeed, in their
important work on the rotational properties of sdB stars, \citet{geier12}
 have found that all the objects in their sample of 105
isolated EHB stars have values of $V_{\rm rot} \sin i$ less than 10 km
s$^{-1}$.\footnote{We note that the much cooler stars with 20,000 K
  $\gta$ \teff $\gta$ 11,500 K belonging to the Blue Horizontal Branch in
  globular clusters are also known to rotate slowly with values of
  $V_{\rm rot} \sin i$ less than about 8 km s$^{-1}$ in the sample of
  \citet{behr2003} and less than 12 km s$^{-1}$ in the sample of
  \citet{recio2002}.} Their distribution of projected rotational 
velocities is 
consistent with a mean value of 7.7 km~s$^{-1}$ for EHB stars. This
figure implies that there must be a significant spin-up during the post-EHB
phase of evolution. As pointed out by \citet{lat17}, in the
absence of significant mass loss and accretion of matter, conservation
of angular momentum in such isolated stars appears to be the most likely
mechanism to account for this spin-up. Indeed, the H-sdOs of interest
here are more compact, with log $g$ $\simeq$ 6.0, than a typical sdB
progenitor, rather characterized by a value of log $g$ $\simeq$ 5.7
(see, e.g., \citealt{heb09}). This translates into a radius ratio of $R_{\rm
  sdB}/R_{\rm H-sdO} \simeq 1.4$. For a {\sl uniform-density} sphere
rotating rigidly, it is well known from classical mechanics that the
equatorial speed $V_{\rm rot}$ scales as $1/R$ if angular momentum is
conserved. Hence, in this rough approximation of constant density, the
H-sdO would rotate some 1.4 times faster than the sdB progenitor. This
falls quite short of the needed factor to account for the values of 25
to 30 km~s$^{-1}$, but, of course, real stars are far from being
uniform-density bodies. Here, we reexamine the question of angular
momentum conservation with the help of realistic evolutionary models. 

\section{Conservation of angular momentum}

\begin{figure}[t]
\begin{center}
\resizebox{\hsize}{!}{\includegraphics{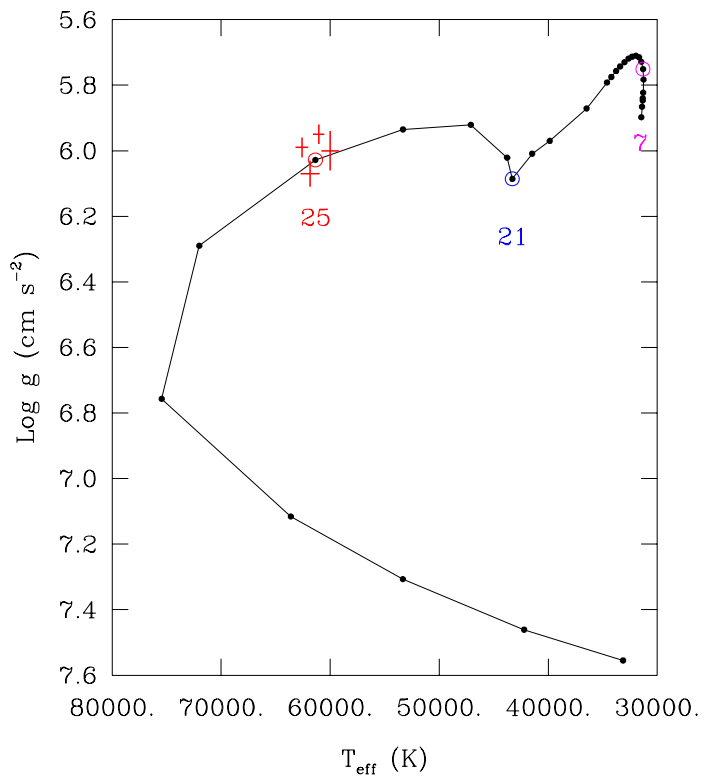}}
\caption{Evolutionary track that passes through the region
of the surface gravity-effective temperature plane where the four sdO
stars of interest are located. Those are depicted by red crosses. Each
available model along the track is indicated by a small black dot. Model
\#7, indicated by a large open magenta circle, is representative of stars
in the EHB phase of evolution and is taken as a reference. Model \#21,
designated by a large open blue circle, indicates the end of the He-core
burning phase. Model \#25, shown by a large open red circle, is
representative of the post-EHB stars of interest.}
\label{track}
\end{center}
\end{figure}

We start by recalling some basic concepts of classical mechanics. 
Consider a sphere rotating uniformly (like a rigid body) about its axis
of symmetry. Its total angular momentum $L$ is given by
\begin{equation}
L = I\omega = \frac{IV}{R}
\end{equation}
where $I$ is the moment of inertia of the sphere with respect
to its symmetry axis, $\omega$ is the angular velocity, $V$ is the
equatorial speed, and $R$ is the radius. Assuming that angular momentum
is conserved, and considering the ``evolution'' of the sphere from
configuration \#1 to configuration \#2, we find that
\begin{equation}
\frac{I_1V_1}{R_1} = \frac{I_2V_2}{R_2}.
\end{equation}

For a {\sl uniform-density} sphere, it is well known from standard
textbooks in classical mechanics that
\begin{equation}
I = {\frac{2}{5}}MR^2
\end{equation}
where $M$ is the mass of the sphere. In that particular case, the
ratio of the equatorial speed in the two configurations becomes,
\begin{equation}
\frac{V_2}{V_1} = {\frac{M_1R_1^2}{M_2R_2^2}}{\frac{R_2}{R_1}} = 
\frac{R_1}{R_2}
\end{equation}
since, by hypothesis, $M_1 = M_2$, i.e., the sphere does not
lose nor accrete matter. In that case, the equatorial speed scales as $V
\propto 1/R$ as mentioned above.

For the more general case of a {\sl non-uniform-density} sphere, we are
left with
\begin{equation}
\frac{V_2}{V_1} = \frac{I_1R_2}{I_2R_1}
\end{equation}
where $I$ must be evaluated numerically according to the following
relation, again available from classical mechanics, 
\begin{equation}
I = \frac{8\pi}{3}{\int_{0}^{R}\rho(r)r^{4}dr}
\end{equation}
where $\rho$ is the density and $r$ is the radial coordinate.

We apply these principles to the case of slowly-rotating stars, i.e.,
with rotation slow enough for Coriolis and centrifugal forces to be
negligible. As indicated above, isolated EHB stars show very little
rotational broadening as a class. This means, at
the very least, that their surface layers rotate very slowly
indeed. Also, from asteroseismology, we know that the internal rotation
profile of at least two pulsating EHB stars \citep{groo08,char08}
is consistent with solid-body rotation over the 
range of depth that can be probed with pressure modes. In
the rest of this paper, and in the absence of other information, we
assume that rigid-body rotation, on which is based the above equations, 
is a good approximation for the stars of interest.

\section{Application to evolutionary models}

\begin{figure}[t]
\begin{center}
\resizebox{\hsize}{!}{\includegraphics{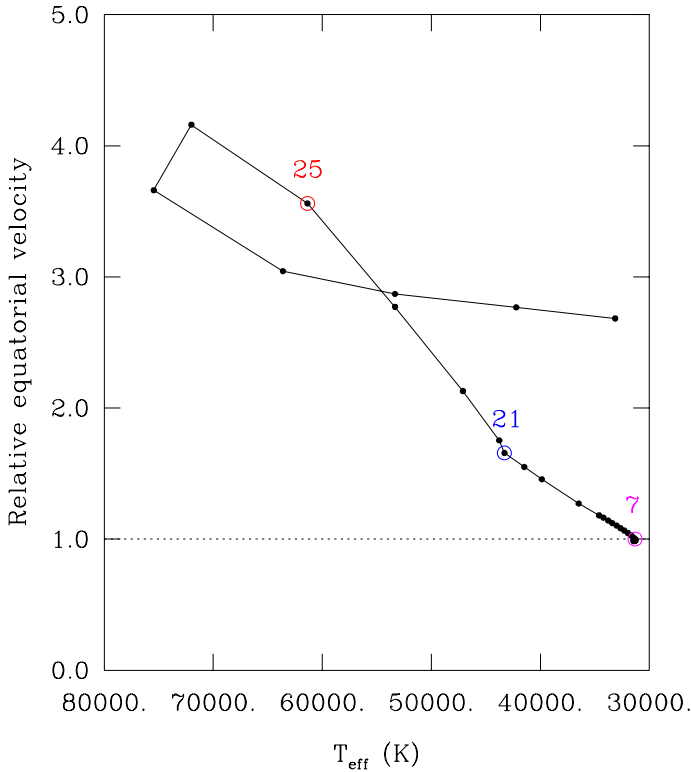}}
\caption{Variation of the equatorial velocity along our
representative track under the assumptions that angular momentum is
conserved during the evolution and that the star model rotates
rigidly. The format is similar to that of Fig. 1. 
Plotted is the ratio of the equatorial velocity with respect to that of
the reference EHB model, Model \#7, as a function of effective
temperature. Model \#25 is characterized by an equatorial velocity that
is some 3.6 times larger than that of the reference EHB model. }
\label{f2}
\end{center}
\end{figure}

\begin{figure}
\begin{center}
\resizebox{\hsize}{!}{\includegraphics{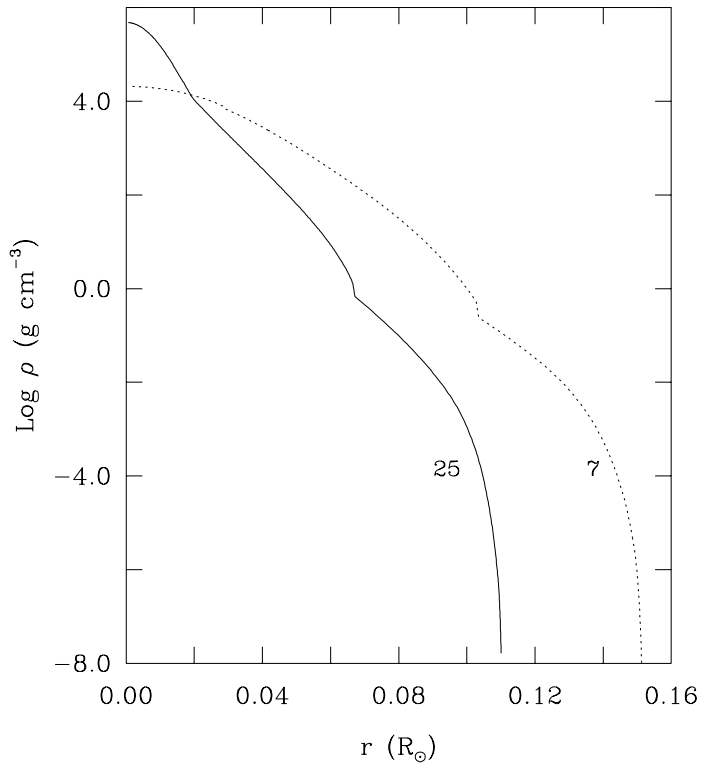}}
\caption{ Density profile in terms of the radial coordinate
(in units of total solar radius) for Model \#7 (dotted curve) compared
to that of Model \#25 (solid curve). The latter has a total radius some
1.4 times smaller than the former, while its central density is about
23.2 times higher. It is this central condensation of matter that is
mostly responsible for the decrease of the moment of inertia of Model
\#25 by a factor of about 4.9 compared to that of Model \#7.}
\label{f3}
\end{center}
\end{figure}

Figure \ref{track} shows the evolutionary track that we picked for our
demonstration in the Kiel diagram. Each small black dot depicts one of
the 31 models that are available along the track, starting with one
model on the Zero Age Extreme Horizontal Branch (ZAEHB) characterized by
\teff = 31\,462 K and log $g$ = 5.898 (the dot at the beginning of the
sequence in the upper right corner). This track is part of a set of seven
evolutionary sequences that have been described at length in Charpinet
et al. (\citeyear{char00,char02}) in their detailed study of the adiabatic pulsation
properties of hot subdwarf stars. They were computed by Ben Dorman
explicitly for the pulsation study of Charpinet et al., based on the
same physics as that described in \citet{dor93}. 

In their post-EHB phase, all of the seven tracks go through basically the
same region where the four H-sdOs of \citet{lat17} are found
(the four crosses in Fig. \ref{track}). The one track we picked, however, has the added
quality that one of its models (Model \#25 indicated by the large open
red circle) happens to fall right within the immediate region in the log
$g$-\teff\ diagram where the rotating H-sdOs are found. Due to the 
relatively poor resolution of the sequences in the post-EHB phase, this
is not the case for the other tracks, and interpolation would be needed
for our comparison work. We thus adopt Model \#25 of the present
sequence as our representative evolutionary model of a hot, H-rich
post-EHB star.

The track illustrated in Fig. \ref{track} belongs to a sequence defined by a
core mass of 0.4758~\Msun\ and an envelope mass of 0.0002~\Msun\ (for a
total mass of $M_*$ = 0.476 \Msun). From this sequence, we also pick a
representative model on the EHB, Model \# 7, chosen somewhat arbitrarily
(another choice of EHB model would have had no significant impact), and
indicated by a large black circle in the plot. This EHB model is burning
He in its core. A period of time of $6.19 \times 10^{7}$ yr has elapsed
since its beginning on the ZAEHB, and it is characterized by \teff =
31\,310 K and log $g$ = 5.751. For its part, Model \#25 has an age of
$1.29 \times 10^{8}$ yr since the ZAEHB and shines through He-shell
burning (which started just after Model \#21 along the track). It is
characterized by \teff = 61\,358 K and log $g$ = 6.028. In comparison,
the average values of the atmospheric parameters of the four H-sdOs of
\citet{lat17} are $<$\teff$>$ = 61\,400 K and $<$log $g$ $>$ = 6.0. 

\begin{figure}
\begin{center}
\resizebox{\hsize}{!}{\includegraphics{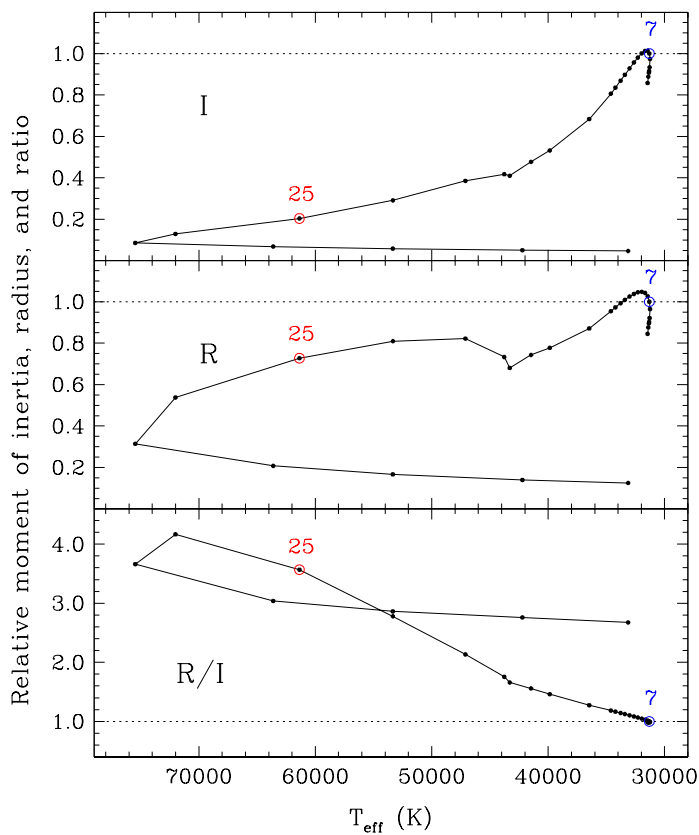}}
\caption{Variation of the relative moment of inertia (top panel), of the
  relative total radius (middle panel), and of the relative ratio of the
  radius-to-the-moment-of-inertia (lower panel) along our representative
  track. The format is similar to that of Fig. \ref{track}. These
  quantities are plotted with respect to those of the reference
  EHB model, Model \#7, as a function of effective temperature. The
  figure reveals that the ratio of $R$/$I$ is maximum for Model \#26.  }
\label{f4}
\end{center}
\end{figure}

We computed the moment of inertia of each of the 31 retained equilibrium
structures along the representative track shown in Fig. \ref{track} with the help
of equation (6). We next computed the ratio of equatorial speeds as given
by equation (5) using Model \# 7 as a reference. The results of these
operations are presented in Fig. \ref{f2}. Of central interest, the figure
shows that, by the time the model has evolved from the EHB to the region
of the log $g$-\teff~diagram where the four H-sdOs are found, 
the equatorial speed has increased by a factor of
$\sim$3.6. Taking the typical value of 7.7 km~s$^{-1}$ obtained by
\citet{geier12} for isolated EHB stars, this translates into a
representative value of $V_{\rm rot} \sin i$ $\simeq$ 27.7 km~s$^{-1}$
for the post-EHB stars near Model \#25 in that diagram
(\teff~$\simeq$ 61\,400 K and log $g$ $\simeq$ 6.0). This corresponds
perfectly to the values of 25 to 30 km~s$^{-1}$ actually inferred by
\citet{lat17} for these stars.

To complement this discussion, we show in Fig. \ref{f3} a comparison of the
density profile of Model \#25 with that of Model \#7. While the radius
of the former is some 1.4 times smaller than that of the latter, its
central density is some 23.2 times higher. We thus find that it is this
phenomenon of central condensation produced by post-EHB evolution that
is mostly responsible for the decrease of the moment of inertia and the
associated spin-up of the model, not the decrease of the radius per
se.

Finally, we plot in Fig. \ref{f4} the individual values of the moment of
inertia $I$ (top panel), of the radius $R$ (middle panel), and of the
ratio $R/I$ (lower panel) for each of the 31 models belonging to the
evolutionary sequence of interest. As above, these three quantities are
normalized to those of Model \#7. Given that the equatorial velocity
scales as $R$/$I$ according to equation (5), and taking into account the
relatively complicated behavior of both $R$ and $I$ in the figure, one
can infer why the relative equatorial velocity depicted in Fig. \ref{f2}
goes through a maximum near the hottest models of the sequence. Of
course, the lower panel of Fig. 4 is nothing more than the equivalent of
Fig. 2 above. It is of some interest to notice that on its cooling branch
(models \#27 through \#31) the equatorial velocity of the resulting white
dwarf decreases with passing time as its radius decreases faster than
does its moment of inertia. 
\section{Conclusion}

The recent work of \citet{lat17}, combined with some previous
indications, has led to the realization that isolated post-EHB stars
generally spin significantly faster than their immediate progenitors,
the EHB stars, known to be slow rotators as a class. Using appropriate
evolutionary models, we have shown that conservation of angular
momentum, coupled to the assumption of rigid rotation (used as a
practical working hypothesis), naturally explains the observed increase
of rotational broadening by a factor of 3 to 4 from the EHB to the
post-EHB phase of evolution.

\begin{acknowledgements}
The authors are grateful to the referee, Dr. Santi Cassisi, for useful
suggestions. This work was supported in part by the NSERC
Canada through a research grant awarded to G.F., and in part by a
fellowship for postdoctoral researchers from the Alexander von Humboldt
Foundation awarded to M.L.. G.F. also acknowledges the contribution of
the Canada Research Chair Program.  
\end{acknowledgements}

\bibliographystyle{aa}
\bibliography{referencef34}

\begin{thebibliography}{24}
\expandafter\ifx\csname natexlab\endcsname\relax\def\natexlab#1{#1}\fi

\bibitem[{{Becker} \& {Butler}(1995)}]{beck95a}
{Becker}, S.~R. \& {Butler}, K. 1995, \aap, 294, 215

\bibitem[{{Behr}(2003)}]{behr2003}
{Behr}, B.~B. 2003, \apjs, 149, 67

\bibitem[{{Charpinet} {et~al.}(2000){Charpinet}, {Fontaine}, {Brassard}, \&
  {Dorman}}]{char00}
{Charpinet}, S., {Fontaine}, G., {Brassard}, P., \& {Dorman}, B. 2000, \apjs,
  131, 223

\bibitem[{{Charpinet} {et~al.}(2002){Charpinet}, {Fontaine}, {Brassard}, \&
  {Dorman}}]{char02}
{Charpinet}, S., {Fontaine}, G., {Brassard}, P., \& {Dorman}, B. 2002, \apjs,
  139, 487

\bibitem[{{Charpinet} {et~al.}(2008){Charpinet}, {Van Grootel}, {Reese},
  {Fontaine}, {Green}, {Brassard}, \& {Chayer}}]{char08}
{Charpinet}, S., {Van Grootel}, V., {Reese}, D., {et~al.} 2008, \aap, 489, 377

\bibitem[{{Deleuil} \& {Viton}(1992)}]{del92}
{Deleuil}, M. \& {Viton}, M. 1992, \aap, 263, 190

\bibitem[{{Dorman} {et~al.}(1993){Dorman}, {Rood}, \& {O'Connell}}]{dor93}
{Dorman}, B., {Rood}, R.~T., \& {O'Connell}, R.~W. 1993, \apj, 419, 596

\bibitem[{{Fleig} {et~al.}(2008){Fleig}, {Rauch}, {Werner}, \& {Kruk}}]{fle08}
{Fleig}, J., {Rauch}, T., {Werner}, K., \& {Kruk}, J.~W. 2008, \aap, 492, 565

\bibitem[{{Fontaine} {et~al.}(2014){Fontaine}, {Green}, {Brassard}, {Latour},
  \& {Chayer}}]{font14}
{Fontaine}, G., {Green}, E.~M., {Brassard}, P., {Latour}, M., \& {Chayer}, P.
  2014, in Astronomical Society of the Pacific Conference Series, Vol. 481, 6th
  Meeting on Hot Subdwarf Stars and Related Objects, ed. V.~{Van Grootel},
  E.~M. {Green}, G.~{Fontaine}, \& S.~{Charpinet}, 83

\bibitem[{{Geier} \& {Heber}(2012)}]{geier12}
{Geier}, S. \& {Heber}, U. 2012, \aap, 543, A149

\bibitem[{{Han} {et~al.}(2003){Han}, {Podsiadlowski}, {Maxted}, \&
  {Marsh}}]{han03}
{Han}, Z., {Podsiadlowski}, P., {Maxted}, P.~F.~L., \& {Marsh}, T.~R. 2003,
  \mnras, 341, 669

\bibitem[{{Heber}(2009)}]{heb09}
{Heber}, U. 2009, \araa, 47, 211

\bibitem[{{Heber}(2016)}]{heb16}
{Heber}, U. 2016, \pasp, 128, 082001

\bibitem[{{Herbig}(1999)}]{her99}
{Herbig}, G.~H. 1999, \pasp, 111, 1144

\bibitem[{{Klepp} \& {Rauch}(2011)}]{klepp11}
{Klepp}, S. \& {Rauch}, T. 2011, \aap, 531, L7

\bibitem[{{Latour} {et~al.}(2018){Latour}, {Chayer}, {Green}, {Irrgang}, \&
  {Fontaine}}]{lat17}
{Latour}, M., {Chayer}, P., {Green}, E.~M., {Irrgang}, A., \& {Fontaine}, G.
  2018, \aap, 609, A89

\bibitem[{{Latour} {et~al.}(2013){Latour}, {Fontaine}, {Chayer}, \&
  {Brassard}}]{lat13}
{Latour}, M., {Fontaine}, G., {Chayer}, P., \& {Brassard}, P. 2013, \apj, 773,
  84

\bibitem[{{N{\'e}meth} {et~al.}(2012){N{\'e}meth}, {Kawka}, \&
  {Vennes}}]{nem12}
{N{\'e}meth}, P., {Kawka}, A., \& {Vennes}, S. 2012, \mnras, 427, 2180

\bibitem[{{Rauch} {et~al.}(2014){Rauch}, {Rudkowski}, {Kampka}, {Werner},
  {Kruk}, \& {Moehler}}]{rauch14}
{Rauch}, T., {Rudkowski}, A., {Kampka}, D., {et~al.} 2014, \aap, 566, A3

\bibitem[{{Recio-Blanco} {et~al.}(2002){Recio-Blanco}, {Piotto}, {Aparicio}, \&
  {Renzini}}]{recio2002}
{Recio-Blanco}, A., {Piotto}, G., {Aparicio}, A., \& {Renzini}, A. 2002, \apjl,
  572, L71

\bibitem[{{Ringat} \& {Rauch}(2012)}]{rin12}
{Ringat}, E. \& {Rauch}, T. 2012, in Astronomical Society of the Pacific
  Conference Series, Vol. 452, Fifth Meeting on Hot Subdwarf Stars and Related
  Objects, ed. D.~{Kilkenny}, C.~S. {Jeffery}, \& C.~{Koen}, 71

\bibitem[{{Stroeer} {et~al.}(2007){Stroeer}, {Heber}, {Lisker}, {Napiwotzki},
  {Dreizler}, {Christlieb}, \& {Reimers}}]{stro07}
{Stroeer}, A., {Heber}, U., {Lisker}, T., {et~al.} 2007, \aap, 462, 269

\bibitem[{{Van Grootel} {et~al.}(2008){Van Grootel}, {Charpinet}, {Fontaine},
  \& {Brassard}}]{groo08}
{Van Grootel}, V., {Charpinet}, S., {Fontaine}, G., \& {Brassard}, P. 2008,
  \aap, 483, 875

\bibitem[{{Viton} {et~al.}(1988){Viton}, {Burgarella}, {Cassatella}, \&
  {Prevot}}]{vit88}
{Viton}, M., {Burgarella}, D., {Cassatella}, A., \& {Prevot}, L. 1988, \aap,
  205, 147

\end{thebibliography}

\end{document}